# Microwave Power Spectra of Stimulated Phonon Emission and Spatio-Temporal Structures in an Optical-Wavelengths Acoustic Laser (Paramagnetic Phaser)

D.N. Makovetskii

(*Instutute for Radio-Physics and Electronics of Nat. Acad. Sci. of Ukraine, Kharkov*)

**Abstract.** A problem of self-organized motions in solid-state nonequilibrium media has been studied experimentally using methods of quantum acoustics. Generalized Poincare cross-sections of microwave power spectra (MPS) have been obtained in an optical-wavelengths acoustic laser (paramagnetic phaser) based on ruby crystal. Considerable narrowing of MPS and their autowave-like superslow motion have been observed under conditions of periodical pump modulation beyond the region of the phaser relaxation resonance. Some preliminar results of this work were published in: Solid State Communications. - 1994. - V.90, No.8. - P.501-505.

The investigation of regular and chaotic dynamical processes in quantum devices of the microwave (MW) and high-frequency (HF) ranges arouses great interest because of an extremely low level of internal noise (spontaneous emission). This permits such dissipative system to be considered as deterministic in all practically accessible ranges of their control parameters. Quantum MW frequency standards (maser MW generators [1, 2]), quantum paramagnetic MW amplifiers with nonlinear pumping resonators [3-5], radiofrequency masers based on the nuclear magnetic resonance [6, 7], as well as optical-wavelengths acoustical lasers (paramagnetic phasers), i.e. quantum paramagnetic amplifiers and generators of MW phonons [8-11], belong to these systems. Unlike the first group of the devices [1-7], the phasers operate due to stimulated emission (SE) of acoustic waves (called also hypersound waves), but not electromagnetic waves. Inasmuch as the propagation velocity of the hypersound in solid state is about by five orders of magnitude less than the electromagnetic wave speed, the wavelength of the phaser radiation is $\lambda \leq 1$ μm, i.e. it is of the order of the visible light wave-length.

Therefore, contrary to the electromagnetic masers, the laser-like quantum acoustic devices (phaser generators) usually operate in a multimode regime and are quite similar to the solid state class "B" lasers [12] in their dynamic characteristics. Yet a relative level of the spontaneous emission in the phaser generator is by 12-15 orders of magnitude less than in the lasers due to the cubic dependence of the spontaneous emission intensity upon the frequency. Thus, the difficulties over the elimination of the multiplicative noise, which complicate experimental work and interpretation of obtained data in the laser systems [13, 14], are not important to the phaser. In this context, the phaser is, in fact, a strictly deterministic model of the class "B" laser. This allows the concept of quantum radiophysics and optical range electronics to be used to set up experiments in a MW range for investigating laser-type phenomena (both regular and deterministic chaotic), which are not distorted by the influence of the noncontrollable and noneliminable internal sources of quantum noise.

In this paper, the MW power spectra (MPS) have been investigated experimentally for the longitudinal phonons SE in the ruby phaser generator in autonomous and nonautonomous regimes. The nonautonomous regime of generation was provided by a periodical low-frequency modulation of pumping of the active medium $Cr^{3+}:Al_2O_3$ in the acoustic Fabry-Perot resonator (AFPR) with Q-factor $Q_c \approx 10^6$. The experiments have been carried out at a frequency of the electromagnetic pumping $\Omega_p = 23$ GHz, the static magnetic field $H \approx 4$ kOe, modulation frequencies $\omega_m = 5 \div 500$ Hz and at a temperature 1.8 K.

An essential feature of the experiments with the panoramic MPS (when investigating a large-scale spectrum structure) was the use of the stroboscopic method. That made it possible to construct the generalized Poicare cross-sections for the multi-attractor dissipative system under study. The gist of the method consists in a short-time ($\Delta t \ll 2\pi/\omega_m$) multiple periodical MPS strobing on the screen of an oscillograph turned on at the output of a heterodyne spectrum-analyzer with a period $\tau_{strob} = \tau_m \equiv 2\pi/\omega_m$. In this case, the time of the registration of the MPS stroboscopic sets, which is defined by photo-exposure duration, was $\tau_{reg} = n_{ex}\tau_{strob}$, where $n_{ex} \gg 1$.

Fig. 1 shows the generalized Poicare cross-sections for the panoramic MPS (the frequency interval is about 8 MHz) whose center frequency is $\Omega_S^{(c)} \approx 9.1$ GHz and the intermode distance is 310 KHz. The left-side oscillogram in Fig. 1 has been produced in an autonomous SE-regime (when there is no pumping modulation): The width of the Poicare cross-sections $\Delta\Omega_S$ along the abscissa axis is ~7 MHz. An insignificant aperiodic automodulation of the SE (spread of dots on the ordinate axis, exceeding the experimental errors) is observed. The middle oscillogram in Fig. 1. has been registered at 100%-modulation of the pumping with a frequency $\omega_m = 94$ Hz, i.e. over the range of the relaxation resonance of the dissipative phaser system (the ν-resonance) [9-11], which responds to a nonlinear phaser excitation at a frequency of intrinsic oscillations of the phonon intermode intensity $J_\Sigma$. On the middle oscillogram, we can clearly see that a destruction occurs in the near-laminar motion which is exhibited at the autonomous SE (the left-side oscillogram); yet $\Delta\Omega_S$ remains practically unchanged (to be more exact, it slightly increases).

It should be emphasized that this destruction is not complete because the mode SE structure is retained, as is shown by the middle oscillogram in Fig. 1.The dispersion of the Poincare cross-section, $J_P$, is large on the ordinate axis only, but it is not noticeable on the abscissa axis in the given scale (as we will show later that it amounts to units of kilohertz). This result correlates, to a certain degree, with the results of Weiss et al. [15, 16]. In [15, 16] it has been shown that the development of a hydrodynamic type large-scale turbulence of the SE in a nonequilibrium dissipative laser-type system is possible only under a certain condition. For the phaser, it should take the form $\tau_C \gg \tau_1$, where $\tau_1$ is the time of the spin longitudinal relaxation; $\tau_C$ is the life time of the phonons in the AFPR in the absence of pumping and beyond magnetic resonance region of the system.

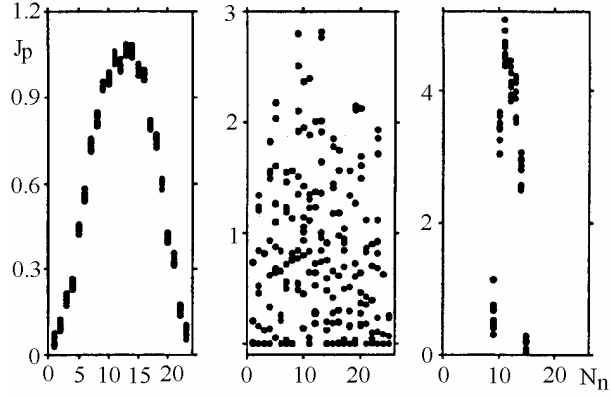

Fig. 1 The generalized Poicare cross-sections for MPS at various regimes: the autonomous regime, i.e. in the absence of the pumping modulation (*the left-side oscillogram*); the pumping modulated in the ν-resonance range (*the middle oscillogram*); and the pumping modulated at the top of the λ-resonance (*the right-side oscillogram*).

In our system this inequality is opposite, therefore, the large-scale mode structure of the SE is retained even if deterministic chaos has developed both for each individual mode (the middle oscillogram) and for the phonon supermode as a whole (see [9-11]). Note that the mode configuration of the panoramic MPS retains its large-scale structure, since the intermode MPS structure is rather complicated even in the autonomous regime [8].

On the other hand, the correlations $\tau_2 \ll \tau_1, \tau_C$ are met in our experiments, where $\tau_2$ is the time of the transverse relaxation of the active medium $Cr^{3+}:Al_2O_3$. Besides the ν-resonance, it means, other nonlinear resonances (λ-resonances [9-11]) can be exited which qualitatively differ from the ν-resonance. If the ν-resonance is less sensitive to the SE spectral structure and suppresses the manifestation of an intermode energy exchange, the whole series of the λ-resonances is formed due to a substantial contribution of processes of the energy redistribution between the spin and phonon degrees of freedom in the phaser active medium subject to an external periodic perturbation. Therefore, it would be natural to expect that the large-scale MPS structure has to undergo visible qualitative changes. That was confirmed by the experiment.

The right-side oscillogram in Fig. 1 was observed at $\omega_m = \omega_\lambda = 9.79$ Hz, where $\omega_\lambda$ is the frequency of the basic λ-resonance [9-11]. It is clearly seen that in this case the width of the phonon MPS decreases by a factor of nearly four as compared with the widths of both autonomous and nonautonomous MPS for $\omega_m = 94$ Hz. At the same time, the chaotic automodulation in the λ-resonances region is much less than in the ν-resonance region. Moreover, in contrast to the ν-resonance, when the existence of the pseudoperiods $\tau_M = 2\pi M / \omega_m$ (*M*=1÷5) is typical for $J_\Sigma(t)$ in the structure of the chaotic automodulation [17], no signs of such pseudoperiods have been revealed at the top of the λ-resonances. Notice that in the present work the measurements of $J_\Sigma(t)$ have been made exactly under the same conditions as in [17], with the exception of values of $\omega_m$. In other words, the discovered MPS renormalization under the influence of the resonant low-frequency perturbation of pumping ($\omega_m = \omega_\lambda$) results from the considerable change in both the topology of the phase space and the metric characteristics of the system.

If the frequency detuning $\Delta_\lambda \equiv \omega_m - \omega_\lambda$ is introduced, the MPS spectra were found to undergo regular self-detunings with a period $\tau_p^{(\lambda)}$. The latter is not commensurable with $\tau_m$ and changes, at least, by five or six orders of magnitude with $\omega_m$ being scanned within the limits of less than 1 Hz in the vicinity of the top of the λ-resonances or some of its highest harmonics. The process of the MPS self-detunings consists in a consecutive displacement of the region of the generating modes localization along the frequency axis in the same direction. At this point the localization of each AFPR mode remains unchanged (let alone the fine effects associated with the internal mode structure). In other words, the consecutive firing of new and new AFPR modes takes place on the one side of the power spectrum of the phonon generation. But on the other side of this spectrum, extinguishing of approximately the same number of modes occurs – up to a break of the generation at some finishing part of the spectrum axis. Afterwards, the process of the global self-detuningis repeated again (from the same starting part of the microwave AFPR modes with the exactly same period).

On the screen of the MW spectroanalyzer, this unexpected effect is visually perceived in the form of a laminar "displacement" along the axis of the frequency of some mode cluster, whose composition varies with time, but the width and form remain much the same. Outwardly, this resembles autowave motion, particularly, dynamics of a burning front propagation [18]. However, it should be kept in mind that all, what is seen on the screen, takes place in a spectrum "space", and concrete spin-phonon structures, being formed in the AFPR, can be evolved in a more complicated way than common autowaves of the real physical spatial continuum [18]. Therefore, when discussing the structures (autostructures) in the phaser system, we will imply any stable spatio-temporal formations having one or another degree of an internal organization.

Let us turn to a concrete description of a set of the experimental data on the phonon SE in the regions where the above-mentioned large-scale laminar MPS self-detunings are exhibited in the nonautonomous phaser. The typical sequences of the SE spectra in the region of the fundamental λ-resonance (of the lowest frequency) are shown in Fig. 2. The spectral sequences A-F are registered experimentally at $\omega_m / 2\pi = 9.5$ Hz (the left-side group of the spectra) and 10.1 Hz (the right-side group); the interval between registration instants in each group was about 2 s. As one can see in Fig. 2, the process of the self-detunings in the power spectrum of the induced phonon field occurs against the background of small-scale irregular pulsations of the intensity of the each mode (as well of insignificant motions of the generating modes along the frequency axis, which are not observed on the panoramic spectrograms in Fig. 2). However, on the whole, the mode structure of the SE corresponds to the initial set of the modes of the longitudinal hypersound in the passive AFPR as in the case of the ν-resonance.

Let us denote by $\Omega_B^{(i)}$ the central frequencies of each mode SE clusters, i.e. instantaneous central frequencies of each moving part of the MPS (there are no more than two of the above parts in Fig. 2). An essential point is that the sign of the derivative $d\Omega_B^{(i)}/dt$ is determined by the sign of detuning of the external-force frequency $\omega_m$ relative to the λ-resonance fundamental frequency $\omega_\lambda$, namely $sgn\left[d\Omega_B^{(i)}/dt\right] = -sgn\,\Delta_\lambda$. As $|\Delta_\lambda|$ approaches to zero, the magnitude $\tau_p^{(\lambda)}$ increases up to very large values. The direct measurements have demonstrated that the lower limit of $\omega_p^{(\lambda)} \equiv 2\pi/\tau_p^{(\lambda)}$ is no more than $10^{-4}$ Hz (one self-detuning period for ≈3 hours).

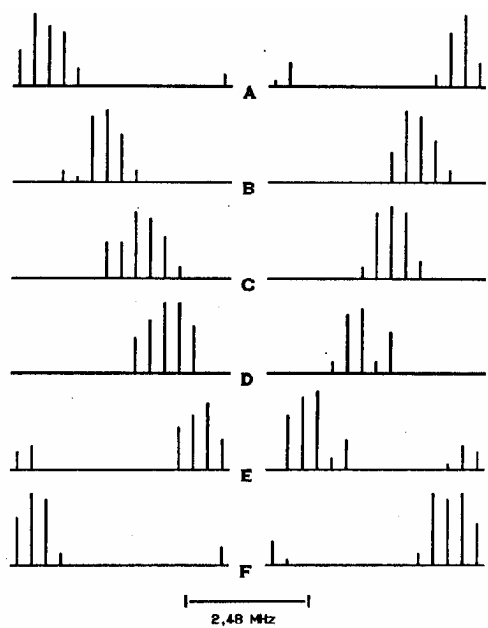

Fig. 2 The MPS large-scale periodic self-detunings at negative (*the left-side spectra*) and positive (*the right-side spectra*) frequency modulation detuning relative to the λ-resonance top.

For $|\Delta_\lambda| \to 0$, a still lesser estimate of the lower limit $\omega_p^{(\lambda)} \leq 10^{-5} \div 10^{-6}$ Hz has been made from the indirect data. These indirect measurements have been carried out by registering the time $\tau_p^{(FD)}$, but not of the value $\tau_p^{(\lambda)}$, where $\tau_p^{(FD)}$ is the time of the minimal displacement of the spectral "autostructure" front (i.e. the front of burning in the spectral space) which can be detected in these experiments. The above-mentioned displacement corresponds to a distance between neighbouring AFPR modes; it is of a discrete character and is about 300 kHz in the spectral space for the AFPR with length 1.7 cm at the longitudinal hypersound velocity of $10^6$ cm/s [11].Since the spectral panorama width reaches 10 MHz, we can obtain the experimental estimate of $\tau_p^{(\lambda)}$, to an accuracy of 30 times greater than for direct measurements $\tau_p^{(\lambda)}$, during the same three hours of experimental working. All the experiments described here were carried our in the superfluid helium; that eliminated obstacles caused by the boiling of a cooling agent. We also note a very high stability of the observed spectral auto-structures with respect to small accidental perturbations of practically all the control parameters of the system (the magnetic field, pumping power, etc.). That enabled us to perform long-time MPS measurements. The stability of MPS formations being observed in the spectral space permits drawing an analogy with auto-waves [18]: both the MPS structures in our experiments and the usual auto-waves do not depend (within certain limits) on the changes in external conditions. In other words, there takes place a kind of self-organized motions in a nonequilibrium dissipative system.

We have every reason to interpret the experimentally discovered phenomena of emergence of the regular spectral auto-structures as a result of the formation of the so called anti-phase states [12, 20] for phonon SE modes when the AFPR exhibits the destabilization of the stationary inter-mode energy exchange at a frequency $\omega_m \approx \omega_\lambda$. Detailed discussion of this interpretation see in our works [9-11].